\documentclass{kapproc} % Computer Modern font calls 

\usepackage{procps}  
 
\usepackage[dvips]{graphicx} 

\upperandlowercase 
 
\kluwerbib % will produce this kind of bibliography entry: 
 
\usepackage{epsfig}

\usepackage[T1]{fontenc}

\begin{document} 
 
\articletitle{The Efficiency of Using Accretion Power of Kerr Black Holes}

%\author{Ioana Du{\textpolhook{t}}an\altaffilmark{1,3} and Peter L. Biermann\altaffilmark{2,3}
%} 
\author{Ioana Du{\c{t}}an\altaffilmark{1,3} and Peter L. Biermann\altaffilmark{2,3}
} 
 
\altaffiltext{1}{University of Bucharest, Faculty of Physics\\
Str. Atomi\c{s}tilor, 405, RO-76900, Bucure\c{s}ti-M\v{a}gurele, Rom\^{a}nia} 
 
\altaffiltext{2}{University of Bonn\\
Regina-Pacis-Weg, 3, D-53113, Bonn, Germany}

\altaffiltext{3}{Max-Planck-Institute for Radioastronomy\\ 
Auf dem H\H{u}gel, 69, D-53121, Bonn, Germany} 
\email{idutan@mpifr-bonn.mpg.de; plbiermann@mpifr-bonn.mpg.de}

\begin{abstract} 
The efficiency of a rapidly spinning Kerr black hole to turn accretion power into
observable power can attain 32 percent for the photon emission from the disk,
as is well known, following the work of Novikov-Page-Thorne. But many accretion disks are now understood to be underluminous ($L<L_{Edd}$), while still putting large amounts of energy into the jet. In this case, the apparent efficiency of jets driven by the innermost accretion disk of a highly rotating Kerr black hole ($a_*=0.999999$) can reach 96 percent.
\end{abstract} 
 
 
\section{Introduction}
The idea of extracting energy from black holes has been proposed by Penrose thirtyfive years ago and followed by a cascade of energy extraction models. Nowadays it is known that from almost all Active Galactic Nuclei (AGNs) which harbour a supermassive ($M>10^5\times M_{\odot}$) Kerr black hole, focused jets of hot gas shoot into space at relativistic speed. "How much energy can these jets get from the black hole or its accretion disk?" is a question which arises by itself.    

\section{Energy extraction from Kerr black holes} 

The Kerr solution (1963) of the Einstein's vacuum field equations describes the gravitational field of a material source at rest having mass and angular momentum. The angular momentum determines a physically significant direction, the axis of symmetry; consequently the field cannot be spherical as in the Schwarzschild solution but axial symmetric. The geometrical units are $G=c=1$, the metric has the signature (-+++), the Greek indices range from 0 to 3, and the asterisk "*" refers to the dimensionless parameters (divided by black hole's mass) throughout the paper.

The metric equation $ds^2=g_{\mu \nu}dx^{\mu}dx^{\nu}$ in the case of a stationary, axial symmetric field takes the form of 
\begin{equation}
ds^2=g_{tt}dt^{2}+2g_{t\phi}dtd\phi +g_{rr}dr^{2}+g_{\theta \theta}d\theta^{2}+g_{\phi \phi}d\phi^{2}
\end{equation} 

The \textit{Boyer-Lindquist coordinates} ($t$, $r$, $\theta$, $\phi$) form a basis of the Kerr spacetime, such that the metric becomes
\begin{eqnarray}
ds^2= -\left(1-\frac{2Mr}{\Sigma}\right)dt^{2}-\frac{4Mar\sin^{2}\theta}{\Sigma}dtd\phi  
       +\frac{\Sigma}{\Delta}dr^{2}+ &\cr +\Sigma d\theta^{2} +\left(r^{2}+a^{2}+\frac{2Ma^{2}r\sin^{2}\theta}{\Sigma}\right)\sin^{2}\theta d\phi^{2} 
\end{eqnarray}  
where $M$ is the mass of the black hole, $a=J/M$ is its angular momentum per
unit mass ($0\leq a\leq M$) and the functions $\Delta$ and $\Sigma$ are
defined by $\Delta =r^{2}-2Mr+a^{2}$ and $\Sigma =r^{2}+a^{2}\cos^{2}\theta $.

Since the metric coefficients in Boyer-Lindquist coordinates are independent of $t$ and $\phi$, both $\zeta^{\mu}=\partial_{t}$ and $\eta^{\mu}=\partial_{\phi}$ are the Killing vectors of the metric, the timelike and axial, respectively, with Boyer-Lindquist components (1,0,0,0) and (0,0,0,1), respectively.

The Kerr solution becomes a singular one, when both $\Delta =0$ 
($g_{rr}\rightarrow \infty $) 
and $\Sigma =0$. There are three possibilities: 
$M^{2}<a^{2}$, $M^{2}=a^{2}$, and $M^{2}>a^{2}$. 
The first two cases are not of physical interest (Carroll 2004) 
and for the third case the equation $\Delta =0$ has two solutions 
$r_{\pm}=M\pm \sqrt{M^{2}-a^{2}}$, where 
both radii are null surfaces which will turn out to be \textit{horizons}, 
the outer $r_{+}$ and the inner $r_{-}$ ones. In the case 
of the Kerr metric the Killing vector is spacelike at the outer horizon, 
except at the north $(\theta=0)$ and south $(\theta=\pi)$ poles, where it is null.

The locus of points where $\zeta^{\mu}\zeta_{\mu}=0$ 
is known as the \textit{Killing horizon} 
(also called the \textit{static limit}), $g_{tt}=0$. 

The region between the Killing horizon and outer horizon is called the ergoregion. Inside the ergoregion no particle can stay at $r, \theta , \phi $ fixed. They are dragged (the Lense-Thirring effect) in the same direction as the black hole rotates, and follow a timelike world line $ds^{2}>0$. These particles have access to negative energy trajectories (Penrose process) which extract energy from the black hole. 

Let us consider the geodesic motion of a particle in the Kerr spacetime. Since the Kerr metric is independent of $t$ and $\phi$ coordinates, there are some conserved quantities for the free motion of a particle, which are associated with $\zeta^{\mu}$ and $\eta^{\mu}$; they are the specific energy ($E^{\dagger}=E/m$) and specific angular momentum ($L^{\dagger}=L/m$) of a particle with mass $m$ orbits on the equatorial plane of a Kerr black hole. But, the circular orbits do not exist for all values of $r$ and also the bound circular orbits are not all stable. There is a peculiar radius, called the radius of the innermost stable orbit $r_{ms}$ from which the orbits of the particles are no longer stable. The value of $r_{ms*}=r_{ms}/M$ depends only on the spin parameter ($a_{*}=a/M$, $-1\leq a_{*}\leq +1$) of the black hole. For more details see (Bekenstein 2004).

\subsubsection{Accretion onto black holes}

The accretion onto Kerr black holes differs from the accretion onto other objects; the absence of the stable circular orbits between the innermost stable orbit and outer horizon implies a rapid flow of matter in black hole, and the presence of the frame-dragging in the ergoregion such that the spin of black hole becomes a source for energy extraction.  

The \textit{Novikov-Page-Thorne model} describes the conversion of accreting mass into outgoing photon energy (Novikov and Thorne 1973, Page and Thorne 1974, Thorne 1974) in the case of a black hole with a classical thin accretion disk lies in the equatorial plane of the hole, such that as the material of the disk spirals inward, releasing gravitational energy, a negligible amount of this energy is stored internally. Almost all energy is radiated away.  

In this model a canonical black hole is the one of spin parameter limited at the value 0.998, by the effect of the photons, which buffer the spin away from the extreme $a_{*}=1$ Kerr value.

The efficiency of converting the accreting mass into radiation measured at infinity, and ignoring small corrections due to capture of photons by the hole, is
\begin{equation}
\eta =1-E^{\dagger}_{ms}=
\left(\begin{array}{l}
\textrm{the specific binding energy of}\\
\textrm{the last stable circular orbit}
\end{array}\right)\\
\end{equation} 
where $E^{\dagger}_{ms}$ is the specific energy of a particle at the innermost stable orbit.

The black hole rotates faster, the higher efficiency of black hole, such that for the canonical black hole the value of the efficiency parameter is around 0.32, lower than 0.42, the maximum value of the efficiency parameter obtained for an extreme Kerr black hole ($a_{*}=1$), by this model. 

\subsubsection{Jets driven by accretion onto Kerr black hole}
Quasars are one of the most energetic astrophysical objects. They live at the centers of galaxies and are manifestation of more general phenomenon of AGNs, which also include Seyfert galaxies, Blazars, Radio galaxies. As is well-known the engines of AGNs are the supermassive black holes and their jets (loud Radio galaxies, Blazars) are associated to the high spin black holes ($a_*\sim 1$). Even though there is no opportunity to watch what is happening when the jets are forming close to the black hole, there are two possible energy sources, the "pure" rotational energy of black hole (Semenov et al. 2004) and the accretion by releasing the gravitational energy of the disk very close to the black hole. Let us concentrate on the last possibility.

The low luminosity of some AGNs (Yuan et al. 2002) could be explained by considering the innermost part of the accretion disk to be non-radiant (Donea and Biermann 1996), such that the only energy-flow out of the innermost disk is along the jets. The magnetic fields of the hole produce a torque on the disk and angular momentum is transported from the accretion flow to the jets through the magnetic stresses. The mass flow rate into the jets is ${\dot{M}}_{jet}={q_m \dot{M}}_D$, where ${\dot{M}}_D$ is the accretion mass onto the hole and $q_m\simeq 0.05$. 
In the following calculations the disk is considered to be nearly Keplerian.

By these considerations, the conservation of the total energy-momentum tensor (Page and Thorne 1974), in the covariant sense $\nabla_{\mu}T^{\mu\nu}=0$, associated with the Killing vectors for the Kerr metric in Boyer-Lindquist coordinates provide the following conservation laws:

\textit{The angular-momentum conservation law}
\begin{equation}
\frac{d}{dr}[(1-q_m){\dot{M}}_{D}L^{\dagger}]=4\pi r(JL^{\dagger}-H)
\end{equation} 
where, $L^{\dagger}$ is the specific angular-momentum of a particle in the disk, J is the flux of energy-flow along the jet, H is the flux of angular momentum (Li 2001) transferred from the black hole to the disk by the magnetic fields, which is given by the magnetic torque produced by the black hole on the disk $T_{HD}= 2\int_{r_{1}}^{r_{2}}2\pi rHdr$, where the factor 2 accounts for the fact that the disk has two surfaces.

\textit{The energy conservation law}
\begin{equation}
\frac{d}{dr}[(1-q_m){\dot{M}}_{D}E^{\dagger}]=4\pi r(JE^{\dagger}-H\Omega_{D})
\end{equation} 
where, $E^{\dagger}$ is the specific energy of a particle in the disk and $\Omega_{D}=1/M(r_{*}^{3/2}+a_{*})$ is the angular velocity of the accretion disk . 

The total energy flow along the jets to infinity is given by
\begin{equation}
E_{jet}=2 \int_{r_1}^{r_2}2\pi JE^{\dagger}rdr=\int_{r_1}^{r_2}[ \frac{d}{dr}((1-q_m){\dot{M}}_{D}E^{\dagger})+4\pi rH\Omega_{D}]dr
\end{equation}
here, the factor 2 from the first right-hand side of the equation accounts for the fact that they are two jets. Considering the jets are formed close to the black hole where the frame dragging process takes place, the inner and outer radii of the above integral could be taken as innermost stable orbit and $r_2=2r_g$, the radius of the ergosphere, respectively, where $r_{g}=GM/c^{2}$ is the gravitational radius. 

\begin{figure}[t] 
\centerline{\includegraphics[width=3.truein]{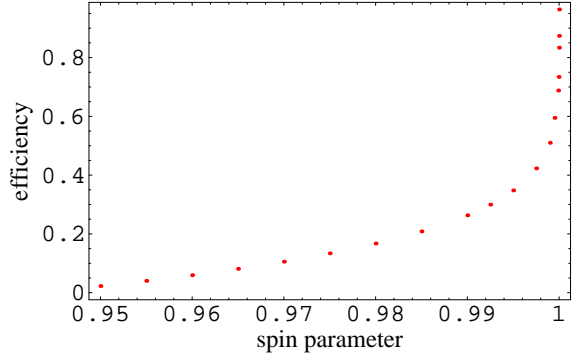}}
\caption{The apparent efficiency $\eta$ of jets driven by accretion disk as function of the spin parameter $a_*$ of a Kerr black hole.}
\label{eff}
\end{figure}

Assuming the disk is perfectly conducting $(Z_{D}\rightarrow 0)$, the flux
of angular momentum transferred from the black hole to the disk by the
magnetic fields is given by $H=\frac{1}{8\pi^3
r}\left(\frac{d\Psi_{HD}}{dr}\right)^{2}\frac{\Omega_{H}-\Omega_{D}}{(dZ_{H}/dr)}$
(Li 2002).
%, where $\Psi_{HD}=\frac{1}{2\pi}\int B^{p}_{D}(dA)_{\theta =\pi
%/2}$ is the magnetic flux through the surface area delimited by the $r_{ms}$
%and $r_2$, for which the only component of magnetic field that counts is the
%poloidal component, $\Omega_{H}=a_{*}/2Mr_{*}=a_{*}/2M(1+\sqrt{1-a_{*}^2})$
%is the angular velocity of a Kerr black hole, and $Z_{H}$ is the surface
%resistance of the hole, $dZ_{H}/dr=2/r$, decreasing with $r$.

The \textit{efficiency of the jets driven} by accretion disk $\eta$ is defined as the ratio of the energy flow along the jets to the rest energy of the accreting mass
\begin{equation}
\eta =\frac{E_{jet}}{\dot{M}_{D}}
\end{equation}
Therefore,
\begin{equation} 
\eta =(1-q_m)[E^{\dagger}(r_{2})-E^{\dagger}(r_{ms})]+\frac{4\pi}{{\dot{M}}_{D}}\int_{r_{ms}}^{r_{2}}H\Omega_{D}rdr
\end{equation}
Here, the second term describes the transfer of energy from the black hole to the disk, and then we assume it can go into jets; and so tapping the energy of the black hole's rotation implies additional energy for the jets, so that the apparent efficiency can be high. 

For an accretion rate $\dot{M}_D$ below the Eddington
($\dot{M}_{Edd}=1.39\times 10^{15}(M/M_{\odot})$ $kg$ $s^{-1}$) rate,
$\dot{m}=\dot{M}_D/\dot{M}_{Edd}\simeq 0.13$, as an example, the value of
efficiency of jets driven by the disk is maximum $\eta_{max}=0.9638$ in the
case of a highly rotating Kerr black hole of a spin parameter $a_*=0.999999$
(see fig. \ref{eff}).

\section{Conclusions}
In this short paper we have tried to describe the energy, mass and angular momentum of a black hole - accretion disk - jet system. We find that the efficiency of such a system putting energy from the accretion into the jet is limited by Thorne's (Thorne 1974) bound of 42 percent, but including the possible transfer of angular momentum and energy from the black hole to the inner disk and then to the jets can temporarily increase the apparent efficiency to near 100 percent. Thus, observed jets might represent such stages of apparent high efficiency.

\begin{acknowledgments} 
I.D. would especially like to thank the Organizers for the financial
support.
I.D. was supported by the
Erasmus/Socrates programme of the EU, the AUGER-project and now by VIHKOS,
as well as the MPIfR. 
\end{acknowledgments}

\end{document}